\documentclass[journal]{IEEEtran} 
\usepackage{epsfig}
\usepackage{epstopdf}
\usepackage{amsmath}
\usepackage{amssymb}

\begin{document}
%
\title{Model-Centric Volumetric Point Cloud Attributes}
\author{
Ricardo L. de Queiroz,\thanks{
R. L. de Queiroz is with the Computer Science Department at Universidade de Brasilia, Brasilia, Brazil, e-mail: queiroz@ieee.org.} 
Camilo Dorea,\thanks{C. Dorea is with the Computer Science Department at Universidade de Brasilia, Brasilia, Brazil, e-mail: camilodorea@unb.br.} 
Davi R. Freitas,\thanks{D. Freitas is with the Post-Graduate Program at the Electrical Engineering Science Department at Universidade de Brasilia, Brasilia, Brazil, e-mail: rabbouni.davi@gmail.com.}
Maja Krivoku\'ca,\thanks{M. Krivoku\'ca is with INRIA, Rennes, France, e-mail: majakri01@gmail.com.} 
and Gustavo P. Sandri.\thanks{G. Sandri is with the Federal Institute of Brasilia, Brasilia, Brazil, e-mail: gustavo.sandri@ieee.org.} 
}

\markboth{Communication}{De Queiroz, 2021}

\maketitle

\begin{abstract}
Point clouds have recently gained interest, especially for real-time applications and for 3D-scanned material, such as is used in autonomous driving, architecture, and engineering, to model real estate for renovation or display.   
Point clouds are associated with geometry information and attributes such as color. 
Be the color unique or direction-dependent (in the case of plenoptic point clouds), it reflects the colors observed by cameras displaced around the object. Hence, not only are the viewing references assumed, but the illumination spectrum and illumination geometry is also implicit. 
We propose a model-centric description of the 3D object, that is independent of the illumination and of the position of the cameras.   
We want to be able to describe the objects themselves such that, at a later stage, 
the rendering of the model may decide where to place illumination, from which it may calculate the image viewed by a given camera.
We want to be able to describe transparent or translucid objects, mirrors, fishbowls, fog and smoke. 
Volumetric clouds may allow us to describe the air, however ``empty'', and introduce air particles, in a manner independent of the viewer position.  
For that, we rely on some eletromagnetic properties to arrive at seven attributes per voxel that would describe the material and its color or transparency. Three attributes are for the transmissivity of each color, three are for the attenuation of each color, and another attribute is for diffuseness.
These attributes give information about the object to the renderer, with whom lies the decision on how to render and depict each object.
\end{abstract}


\section{Introduction}

\begin{figure}[t]
	\centering
	\
	\includegraphics[width=.40\textwidth]{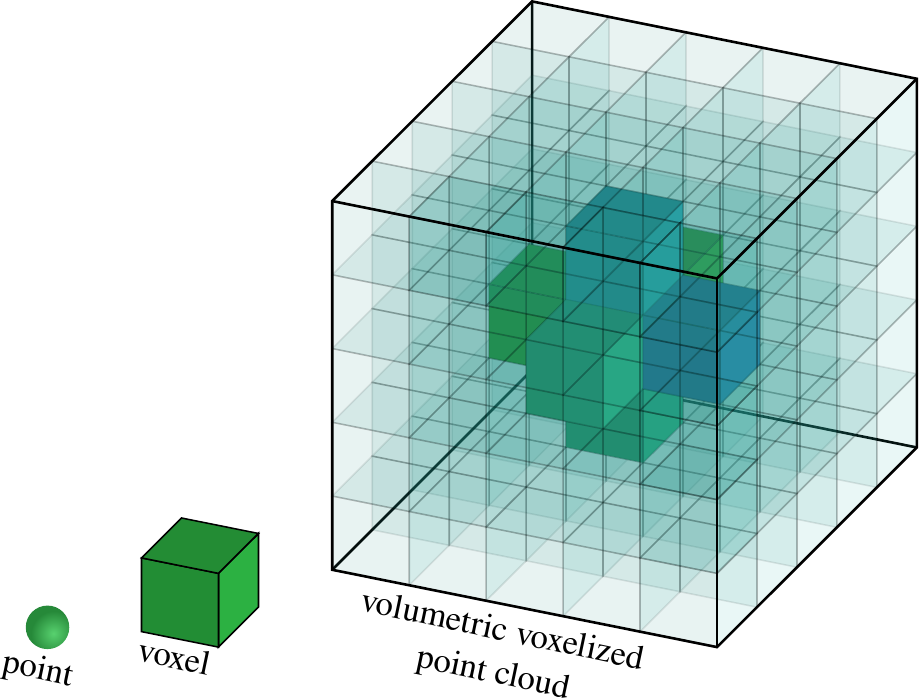} 
	\caption{3D objects can be represented by a cloud of points or a grid of voxels, encompassing occupied or empty voxels. Conceptually, all voxels may receive attributes.}
	\label{fig:volumetricpc}
\end{figure}

Point clouds are becoming more popular today for representing 3D objects and people in scenarios that are captured from the real world, without the additional complexity of deriving meshes or other connected (surface) structures \cite{Gross2007}. 
Points in a cloud are by definition unconnected, and are usually described by a list of points describing the geometry of each point (i.e. the usual $xyz$ coordinates in 3D space) and any attribute the point might have. Examples of attributes are colors, in any color space, reflection intensity, or surface normals.   
There are voxelized point clouds, wherein the space is subdivided into a regular grid of ``\textit{vo}lumetric pi\textit{xels}'' or {\em voxels}.  
One may convert back and forth from voxelized and non-voxelized point clouds, by snapping points to the voxel grid and by creating a point at the center of a voxel. 
Most point clouds, even those that are voxelized, are in fact \textit{surface point clouds}, since they do not represent the interior of objects, just the visible surfaces.  
Another alternative is a \textit{volumetric} point cloud, in which all the voxels inside and on the surface of the 3D object are described, with information on which voxels are occupied or not, perhaps using functions that cover the entire space. 
Fig. \ref{fig:volumetricpc} illustrates the volumetric point cloud case. 

In computer graphics \cite{Foley2013}, describing the model and rendering it are completely different processes. 
The model describes the object, its location and its properties. The rendering is done via ray-tracing or another method, assuming a given model, the illumination, and the camera viewpoint. 
Point clouds, on the other hand, usually mix both concepts, as the voxel color is typically captured by cameras in given positions, subject to a given illumination. 
We propose a point cloud attribute format that is model-centric, illumination-independent, and viewpoint-independent, which would describe the objects themselves independently of the rendering. 

The proposed model-centric concept would provide light independence in describing objects, for example buildings \cite{Matterport}.
The same point cloud could be rendered in daytime, night-time, or even include interior lights. 
Transparent materials could be represented, as well as reflective materials (e.g., brass), or even mirrors.  
We anticipate that scanning transparent objects may be a great challenge. Volumetric objects would need to be represented in this case, instead of just the object surfaces.
Interior materials could also be represented, since visibility is not implied in the proposed point cloud. 
We have developed a set of seven attributes that we believe can be used to reasonably describe the object materials, in terms of their visibility, in a voxelized volumetric point cloud.  
Their reasoning is described next.

\section{Eletromagnetism-based motivation} 

Just a quick word on eletromagnetism, pertaining to transversal eletromagnetic waves in a given medium \cite{Paris69}. 
Light in most cases can be modeled as such and it is of interest to understand what happens to the wave as it changes media.

\subsection{Planar waves}

The wave propagates due to its varying electric and magnetic fields. From Maxwell's equations, variations in the magnetic field induce changes in the electric field and vice-versa. The result is a wave that propagates in space. 
In order to simplify the description, assume a planar wave propagating in one direction, with fields polarized along the perpendicular directions. 
Let ${\bf a}_x$, ${\bf a}_y$ and ${\bf a}_z$ be vectors aligned to the coordinate system and assume the wave is propagating along ${\bf a}_z$ and the electric and magnetic fields are given by 

\begin{equation}
{\bf E} = E_x e^{-\gamma z} {\bf a}_x ,
\end{equation}
\begin{equation}
{\bf H} = \frac{E_x}{\eta} e^{-\gamma z} {\bf a}_y .
\end{equation}

\noindent
where $E_x$ is a time-varying component of the electric field over the $x$ axis and $\eta$ is the impedance of the medium. 
This is a linearly polarized wave. 
Both $\eta$ and $\gamma$ are potentially complex numbers. 
If we express 

\begin{equation}
\gamma = \alpha + j \beta , \footnote{We adopt the $j = \sqrt{-1}$ notation for complex numbers.}
\end{equation}

\noindent
so that 
\begin{equation}
{\bf E} = E_x e^{-\alpha z} e^{-j \beta z} {\bf a}_x ,
\end{equation}

\noindent 
which tells us that $\alpha$ is responsible for attenuating the wave while  $\beta$ is its phase propagation term. 
The same occurs for the magnetic field. 

The wave is assumed as having a frequency $\omega = 2\pi f = 2\pi/\lambda$, where $\omega$ is given in radians per second, $f$ is the frequency in Hertz and $\lambda$ is its wavelength.

\subsection{Light in a medium}

Before we describe the quantities $\gamma$ and $\eta$ in the previous section, keep in mind that a medium is associated with three eletromagnetic quantities. 
There is the electric permittivity $\varepsilon = \varepsilon_r \varepsilon_0$, where $\varepsilon_r$ is the relative permittivity (or dielectric constant) of the medium and $\varepsilon_0$ is the permittivity of the vacuum. 
There is also the magnetic permeability $\mu = \mu_r \mu_0$, where $\mu_r$ is the relative permeability of the medium and $\mu_0$ is the permeability of the vacuum. 
$\sigma$ is the conductivity of the medium.

The wave propagation term is then 

\begin{equation}
\gamma = \sqrt{ j\omega\mu (\sigma + j\omega\varepsilon) } , 
\label{eq:gamma}
\end{equation}

\noindent 
i.e.

\begin{equation}
\alpha = \omega\sqrt{\mu\varepsilon}\sqrt{\frac{1}{2} \left[ \sqrt{1+\left(\frac{\sigma}{\omega\varepsilon}  \right)^2} - 1\right]} , 
\end{equation} 

\begin{equation}
\beta = \omega\sqrt{\mu\varepsilon}\sqrt{\frac{1}{2} \left[ \sqrt{1+\left(\frac{\sigma}{\omega\varepsilon}  \right)^2} + 1\right]} .
\end{equation} 

The medium impedance is 

\begin{equation}
\eta = \sqrt{ \frac{j\omega\mu}{\sigma + j\omega\varepsilon} } . 
\label{eq:impedance}
\end{equation}

\noindent
The wave travels in such a medium with a phase velocity (which, in this case, is a pretty good approximation to the true velocity) given by 

\begin{equation}
v = \frac{1}{\sqrt{\frac{\mu\varepsilon}{2} \left[1 + \sqrt{1+\left(\frac{\sigma}{\omega\varepsilon}  \right)^2}\right]}}
\end{equation}

\subsection{Cases of interest}

The above equations impose some algebraic difficulties, but we are most interested in two cases: the good dielectric and the good conductor.  
For the good dielectric, we assume $\sigma$ to be very small, such that $\alpha \approx 0$, $\beta \approx \omega \sqrt{\mu\varepsilon}$, and the impedance is: 

\begin{equation}
\eta \approx \sqrt{ \frac{\mu}{\varepsilon} } , 
\end{equation}

\noindent 
with equalities for a perfect dielectric material ($\sigma=0$).
For the same conditions, the speed in the medium becomes 

\begin{equation}
v \approx  \frac{1}{\sqrt{\mu\varepsilon} } . 
\end{equation}

A special case of a good dielectric is the open air, which approximates the vacuum, wherein $\mu_r=\varepsilon_r=1$, such that $\eta=\sqrt{\mu_0/\varepsilon_0} \approx 377\ \Omega$. The speed is  $c = 1/\sqrt{\mu_0\varepsilon_0}$, which is the causality speed of the universe or the ``speed of light'', and it is the maximum speed any wave can travel. 

In the case of a good conductor and when $\sigma$ is large, 

\begin{equation}
\eta \approx \sqrt{\frac{\omega\mu}{\sigma}} e^{j\pi/4} 
\ \ \mbox{and} \ \
v  \approx \sqrt{\frac{2\omega}{\mu\sigma}} .
\end{equation}

One last important note is that when there is conductivity, the propagation is attenuated. Any imaginary part of the impedance term causes attenuation. 
A good conductor, thus, does not allow propagation in its core.
In fact the eletromagnetic wave barely penetrates a conductor, as much as what is named the {\em skin depth}, which is the point where the electric field has dropped by a factor of $e^{-1}$. 
The skin depth is $\delta = v / \omega$. For 100 MHz and copper, for example, $\delta = 6.6 \mu m$.

\subsection{Light crossing a media interface}

Assume, then, light as a planar wave crossing an infinite planar interface (perpendicular to normal vector $\bf n$) in between materials 1 and 2. 
Assume the origin to be where the wave hits the planar surface.
Assume that the incident light comes from direction ${\bf n}_i$.
It is well known that some light is reflected at direction ${\bf n}_r$ while some is transmitted (refracted) into medium 2 at direction ${\bf n}_t$.
All four vectors lie in the plane defined by vectors ${\bf n}_i$ and ${\bf n}$.
If all vectors have unit norm,  ${\bf n}_r = 2 ({\bf n}_i \cdot {\bf n})\ {\bf n} - {\bf n}_i$.  
Note that $\theta_1 = {\bf n}_i \cdot {\bf n} = {\bf n}_r \cdot {\bf n}$ is the incidence (and reflectance) angle, while the transmission  (refraction) angle $\theta_2$ is given by Snell's law:

\begin{equation}
\frac{\sin \theta_2} {\sin \theta_1}  = \frac{v_2}{v_1} ,
\label{eq:snell}
\end{equation}

\noindent 
where the index indicates the medium. If both media are good dielectrics, we have  

\begin{equation}
\frac{\sin \theta_2} {\sin \theta_1}  =  \frac{\sqrt{\mu_1\varepsilon_1}}{\sqrt{\mu_2\varepsilon_2}} = 
\frac{\beta_1}{\beta_2} .
\end{equation}

Snell's equation is sometimes expressed in terms of the refraction index which is $c/v$. 
Such a redundant quantity was left out of our discussion.

The refraction (transmission) vector is: 
\begin{equation}
{\bf n}_t = (\sin \theta_2 \cos \theta_1 - \cos \theta_2) {\bf n} - {\bf s}_i .
\end{equation}

\begin{figure}[t]
	\centering
	\
	\includegraphics[width=.30\textwidth]{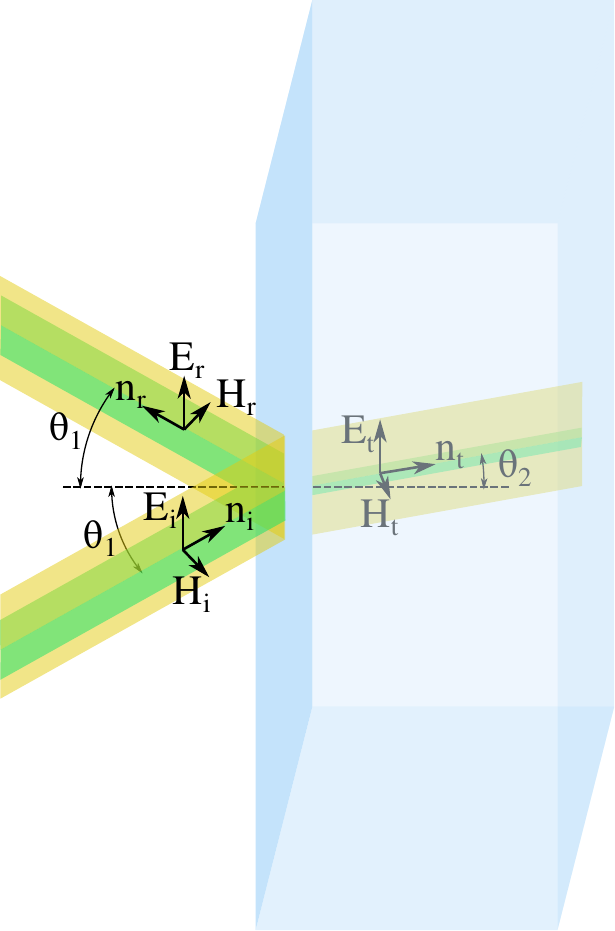} 
	\includegraphics[width=.30\textwidth]{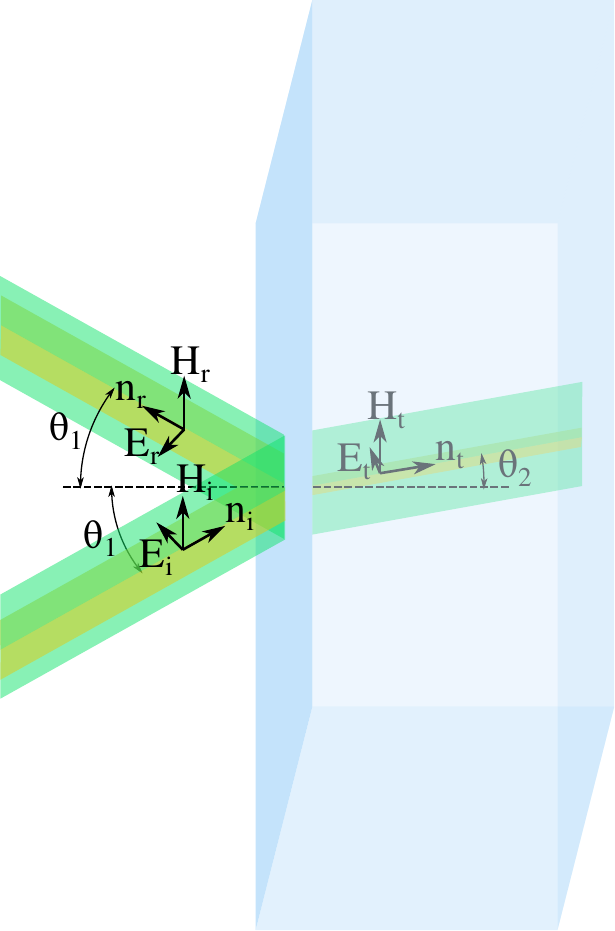} 
	\caption{Wave incidence geometry on a planar interface. Top: $E$ perpendicular to the surface; bottom: $E$ parallel to the surface.}
	\label{fig:planarincidence}
\end{figure}

There are three waves to consider: the incident (fields ${\bf E}_i$ and ${\bf H}_i$), the reflected (fields ${\bf E}_r$ and ${\bf H}_r$), and the transmitted or refracted (fields ${\bf E}_t$ and ${\bf H}_t$), which are depicted in Fig. \ref{fig:planarincidence}.
Since the magnitude of the electric and magnetic fields are related by the impedance, we are concerned with the propagation ratio of the electric field into medium 2, i.e., $T=E_t/E_i$ and the amount of the field that is reflected $\Gamma=E_t/E_i$.
As also depicted in Fig. \ref{fig:planarincidence}, there are two cases to consider, and a more general case which is a combination of both: when the electric field ${\bf E}_i$ is parallel to the surface (${\bf n}\cdot {\bf E}_i = 0$)

\begin{equation}
\Gamma_\perp = \frac{E_r}{E_i} = 
\frac {\eta_2 \cos \theta_1 - \eta_1 \cos \theta_2}{\eta_2 \cos \theta_1 + \eta_1 \cos \theta_2 } , 
\label{eq:fresnel1}
\end{equation}

\begin{equation}
T_\perp = \frac{E_t}{E_i} = 
\frac {2\ \eta_2 \cos \theta_1}{\eta_2 \cos \theta_1 + \eta_1 \cos \theta_2 } ,
\label{eq:fresnel2}
\end{equation}

\noindent
and when the magnetic one is parallel to the surface (${\bf n}\cdot {\bf H}_i = 0$)   

\begin{equation}
\Gamma_\parallel = \frac{E_r}{E_i} = 
\frac {\eta_1 \cos \theta_1 - \eta_2 \cos \theta_2}{\eta_1 \cos \theta_1 + \eta_2 \cos \theta_2 } ,
\label{eq:fresnel3}
\end{equation}

\begin{equation}
T_\parallel = \frac{E_t}{E_i} = 
\frac {2\ \eta_2 \cos \theta_1}{\eta_1 \cos \theta_1 + \eta_2 \cos \theta_2 } .
\label{eq:fresnel4}
\end{equation}

\subsection{Considerations} 

In an imperfect dielectric, a small but non-zero $\sigma$ may cause a small but significant attenuation. 
Attenuation, in general, is caused by ``absorption'', which is dissipation of the energy in the material, being converted into heat.  
The reason is that conductivity causes minor currents which have dissipation losses due to electrical resistance. 

As conductivity increases, the refraction angle decreases. For a good conductor, the field rapidly attenuates but $\theta_2 \approx 0$, i.e., the transmitted wave is refracted along the normal no matter where the incident wave comes from.    

For non-ferromagnetic materials, $\mu_r \approx 1$ and, for perfect dielectric materials, 

\begin{equation}
\eta \approx \sqrt{\frac{\mu_0}{\varepsilon_r \varepsilon_0}} = \frac{1}{\sqrt{\varepsilon_r}} \eta_0  
\label{eq:impedancedielectric}
\end{equation}

\noindent 
and 

\begin{equation}
v \approx \sqrt{\frac{1}{\varepsilon_r \varepsilon_0 \mu_0}} = \frac{1}{\sqrt{\varepsilon_r}} c  
\label{eq:speeddielectric}
\end{equation}

\noindent
where $\eta_0$ is the impedance of vacuum and $c$, again, is the causality speed or the light speed in the vacuum, the maximum speed there is.  
In this case, $1/\sqrt{\varepsilon_r}$ modulates both speed and impedance, which we use to calculate the refraction angle ($\theta_2$) and the transmission ($T$) and reflection ($\Gamma$) coefficients. 

In this case, as the relative permittivity $\varepsilon_r$ drops to 1, we obtain the impedance and speed of vacuum.  
Conversely, for very large permittivity, both impedance and speed drop. 
In the limit, if medium 2 has such large permittivity and medium 1 does not, $\Gamma\rightarrow -1$ and $T\rightarrow 0$, i.e., there is total reflection.

\subsection{Light Beam Diffusion}

The assumption so far  was that of a planar wave incident onto a planar surface separating two uniform media.  
This is never completely true. 
The rugosity of a surface, even at a micron level, may make a light beam appear as having multiple incident angles since the wavelength of visible light is in the order of 0.5 micron.  
This could result in dispersion of both the transmitted and reflected beams. 
This is the reason why conductor materials are not necessarily good mirrors. 
Commercial mirrors are usually made of a conducting material, first in liquid form,  pressed within smooth plates, one of then usually made of glass. 

Another source of light dispersion is the layering of materials, like coatings on surfaces, which cause many reflections and refractions from stacked rough planes to be combined, thus forming diffuse beams. 

Another beam deformity can be caused by heterogeneous materials. 
For example, a perfect dielectric may be polluted with  conductive or high-impedance materials, in a mix of attenuation, reflection and refraction. 
One example is sea water, which can be clean and transparent, or murky and dark, depending on the sediments in suspension.
Similarly, mist or smoke can dampen light propagation in air.

\section{Attributes}

In our use case of a voxelized point cloud, surface refraction and reflection occurs at voxel boundaries. 
A voxel represents a uniform medium, but its boundaries are too large compared to visible light wavelength. 
If a voxel represents a cubic centimeter of space, for example, all material imperfections and rugosity at the millimeter down to micron scale may cause dispersion and diffusion. 

We need measures that are simple, normalized from 0 to 1, for example, without reverting to difficult properties such as permittivity or conductivity. 
Yet, we need to be able to calculate refraction angles, beam dispersal and transparency. 

We want to describe on a per-voxel basis: how much light is reflected and how it refracts; how much light is attenuated within the medium; how light is dispersed when crossing rough surfaces at sub-voxel scales. 
We propose to use 3 parameters, as described below. 

\noindent {\bf Transmissivity -} 
This parameter $P_t$ controls how much light is absorbed into the voxel, and is related to the inverse of the medium eletromagnetic impedance. 
It ranges from 0.0 to 1.0.
When $P_t = 0$, we mean that light is completely propagated into the voxel. 
Conversely, when $P_t = 1$, the light is mostly reflected by the voxel. 

\begin{figure}[t]
	\centering
	\
	\includegraphics[width=.38\textwidth]{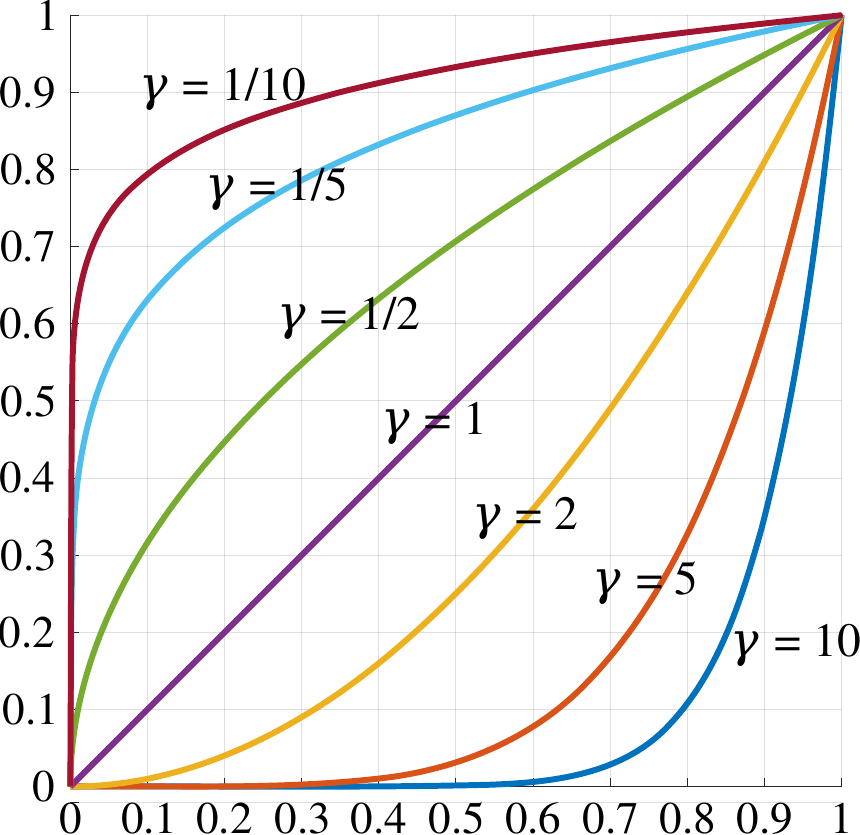}
	\caption{Gamma correction for parameter assignment. }
	\label{fig:gamma}
\end{figure}

One rendering suggestion is to associate $P_t$ with $\varepsilon_r$, for example, setting $P_t = k_t \log (\varepsilon_r)$ for $k_t$ some constant of comfort. 
Since $\varepsilon_r \geq 1$, the idea is that $P_t = 0.0$ would represent the permittivity of the vacuum, while a sufficiently large permitivitty can be made to  correspond to $P_t = 1.0$. 
Further biasing of the correspondence can be made using "gamma" correction, as depicted in Fig. \ref{fig:gamma}, where the typical ``gamma'' parameter is not related to the parameter with the same symbol from the previous Section. 
In the junction between two voxels, from $P_1$ one may find $\varepsilon_r$, more precisely $1/\sqrt{\varepsilon_r}$ of each medium, from where we can refer to Eq. (\ref{eq:speeddielectric}) in order to estimate velocities, and use Snell's equation 
(\ref{eq:snell}) to find refraction angles.  
From the $1/\sqrt{\varepsilon_r}$ of each medium, one can also calculate the media impedance using Eq. (\ref{eq:impedancedielectric}), from which the intensity of light reflected of refracted can be calculated form Fresnel equations in Eqs. (\ref{eq:fresnel1})
through (\ref{eq:fresnel4}).

\noindent {\bf Attenuation -} This parameter $P_a$ could be useful to express the attenuation of the medium. 
It may be related to the conductivity of the medium $\sigma$, but it may also indicate how much of the incoming (refracted) light is to be propagated to the next voxel. 
For example, the quantity $P_a=0.0$ may indicate a conductor with (near) total attenuation. Recall that the ``skin depth'' of a conductor at visible light frequency (in the order of $10^{15}$ rad/s) is incredibly small compared to a voxel size.     
$P_a=1.0$ may indicate that all incoming light to a voxel is passed to the next one. 

\begin{figure}[t]
	\centering
	\
	\includegraphics[width=.48\textwidth]{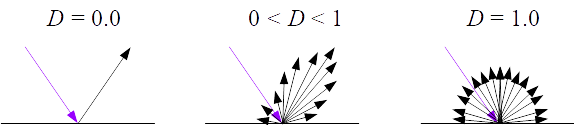}
	\caption{Diffusion parameter relfection examples. When $D=0$, there is perfect specularity and light is reflected without dispersion. As $D$ increases, reflected light becomes more diffuse with some Lambertian reflection. For maximum diffusion ($D=1$), reflection is completely Lambertian with no preferred direction regardless of the incidence geometry. Similar concepts apply to transmitted (refracted) light.   }
	\label{fig:diffusion}
\end{figure}

\noindent {\bf Diffuseness -}  This parameter $D$ controls the diffusion of the incident light caused by the non-uniformity of the interface at sub-voxel dimensions. 
Since it is caused by the rugosity of the interface, the dispersion affects both the reflection and the transmission (refraction). 
In a sense, it is a measure of specularity of the voxel surface. 
If $D=0$, we assume there is no dispersion and light is reflected (refracted) with perfect specularity (transmission).
If $D=1$, the material is assumed perfectly Lambertian and light is reflected (refracted) equally in all directions.
Intermediary values of $D$ assume a progressive increase in dispersion of the main beam and increasing Lambertian properties.   
This is illustrated in Fig. \ref{fig:diffusion}.
Rendering for those intermediary values can use, for example, the dispersion of the Phong \cite{Phong75} or Blinn-Phong \cite{Blinn77} models. 
The Schlick model \cite{Schlick94} assumes multiple layers, and that may be accounted for by imposing diffusion over multiple layers of voxels.

\section{Colors and transparency}
 
Both the impedance and the propagation speed are dependent on $\omega$, the wave frequency. 
For a very small conductivity, the wave across two dielectric materials is much more dependent on the permittivity than on the frequency. 
The permittivity, however, may also depend on the frequency.  
Rather than defining $\varepsilon$ for a spectrum of frequencies, since colors are commonly described in terms of red, green and blue components, we use these three color components. They act like filtered averages of the spectrum, or, in some sense, they are some kind of sampling of the spectrum. 

Thus, in order to convey the colors of the materials, as a reflection or refraction of a white incident color (flat spectrum, or equal red-green-blue components), we propose to convey the parameters $P_a$ and $P_t$ for each component (red, green and blue). 
It is reasonable to assume that the rugosity which causes diffusion has dimensions larger than the wavelengths of either red, green or blue components. 
Hence, $D$ may not need to be represented for each color component. 

As a result, we propose to use seven parameters as attributes to describe the materials for each voxel of a volumetric point cloud. 
The parameters are: $R_t$, $G_t$, $B_t$, $R_a$, $G_a$, $B_a$, and $D$. 
The first three are the $P_t$ (transmissivity) for each color component, with similar notation for the next three in terms of $P_a$ (attenuation). 
In summary, we propose the attributes in Table \ref{tab:proposal}.

\begin{table}[h]
	\centering
	\caption{Proposed attributes per voxel.}
	\begin{tabular}{|l|c|l|c|l|} \hline
		\ & Min. & Meaning & Max. &  Meaning \\ \hline 
		$R_t$ & 0.0 & All red transmitted & 1.0 & All red reflected \\ 
		$G_t$ & 0.0 & All green transmitted & 1.0 & All green reflected \\ 
		$B_t$ & 0.0 & All blue transmitted & 1.0 & All blue reflected  \\ 
		$R_a$ & 0.0 & No red attenuation & 1.0 & All red absorbed \\ 
		$G_a$ & 0.0 & No green attenuation & 1.0 & All green absorbed \\ 
		$B_a$ & 0.0 & No blue attenuation & 1.0 & All blue absorbed \\ 
		$D$ & 0.0 & Perfect specularity  & 1.0 & Lambertian diffusion \\ \hline
	\end{tabular}
	\label{tab:proposal}
\end{table}

In Table \ref{tab:examples} we list examples of attributes for a number of materials of interest. 
All the parameter numbers are guessed examples just to illustrate how the 7 parameters may describe the appearance of the diverse materials. 
These numbers are not of any quantitative interest, just for qualitative illustration. We only used simplifying numbers 0.0, 0.2, 0.5, 0.8, and 1.0 to represent ``nothing'', ``a little'', ``medium'', ``a lot'', and ``all''. 
We used shirts as examples of a Lambertian surface that reflects the light in different amounts to give the color appearance of the shirt.
Note that all shirts have full attenuation (they are not transparent) and maximum diffusion. The more they reflect, the lighter they appear. 
Brass is conductive with large attenuation. 
Brass objects are usually shiny objects, i.e., reflective depending on how polished they are. We assumed brass to be somewhat polished and with a small but significant diffusion. 
Glass is a transparent object. We assumed it not to be perfectly transparent, with $P_t=0.2$. 
We also assumed a small attenuation and a very flat glass without diffusion.  
We included a large diffusion in frosted glass.  
Similarly for water, where we changed the values to include more ``blue'' reflection. 
Air (clear air) has $P_t=P_a=D=0$, i.e., no reflection, no attenuation, no diffusion. 
Smoke or mist may be represented by increasing a little some of those parameters. 
In any case, these parameters are only illustrative, with no empirical evaluation.

\begin{table}[h]
	\centering
	\caption{Example parameters for a few materials.}
	\begin{tabular}{|l|c|c|c|c|c|c|c|} \hline
		Material & $R_t$ & $G_t$ & $B_t$ & $R_a$ & $G_a$ & $B_a$ & $D$ \\ \hline 
		White shirt & 0.8 & 0.8 & 0.8 & 1 & 1 & 1 & 1 \\ \hline
		Dark shirt & 0.2 & 0.2 & 0.2 & 1 & 1 & 1 & 1 \\ \hline
		Red shirt  & 0.8 & 0 & 0 & 1 & 1 & 1 &  1 \\ \hline
		Green shirt & 0 & 0.8 & 0 & 1 & 1 & 1 & 1 \\ \hline
		Blue shirt & 0 & 0 & 0.8 & 1 & 1 & 1 & 1 \\ \hline
		Color shirt & 0.8 & 0.5 & 0.2 & 1 & 1 & 1 & 1 \\ \hline
		Skin & 0.5 & 0.5 & 0.2 & 0.8 & 1 & 1 & 0.8 \\ \hline
		Brass & 0.8 & 0.8 & 0.2 &  1  &  1 & 1 & 0.2 \\ \hline
		Glass & 0.2 & 0.2 & 0.2 &  0.2 & 0.2 & 0.2 & 0 \\ \hline
		Frosted Glass & 0.2 & 0.2 & 0.2 &  0.2 & 0.2 & 0.2 & 0.8 \\ \hline
		Water & 0.2 & 0.2 & 0.5 & 0.2 & 0.2 & 0.2 & 0 \\ \hline
		Mirror & 1 & 1 & 1 & 1 & 1 & 1 & 0 \\ \hline
		Air  & 0 & 0 & 0 & 0 & 0 & 0 & 0 \\ \hline
		Smoke/mist  & 0.5 & 0.5 & 0.5 & 0.2 & 0.2 & 0.2 & 0.5 \\ \hline
	\end{tabular}
	\label{tab:examples}
\end{table}


\section{Conclusions}
\label{sec:conclusions}

The purpose of this communication is to start a conversation on a volumetric-point-cloud model description that is independent of lighting and of camera viewpoints.  
Rendering, or capturing/scanning should be independent of the model description, allowing for example to change lighting. 
One example is to be able to represent a building at day or night, even placing illumination therein.   
All these options should be rendering choices, not an intrinsic part of the point cloud. 

Our next steps for such point clouds are: (i) synthetic generation; (ii) rendering; (iii) compression; and (iv) scanning. 

This proposal is intended for non-luminous materials only. Internal lights, of any luminescence, which may include diverse light bulbs, may be represented under different sets of attributes of the same point cloud. 


\section{Acknowledgment}
\label{sec:acknowledgment}
The authors would like to thank Prof. Leonardo R. A. Xavier, from Universidade de Brasilia, for the help with  eletromagnetism issues.

\end{document}